\documentclass{ws-ijmpe}
\usepackage[super,compress]{cite}
\usepackage{color}
\usepackage{amsmath}
\usepackage{epstopdf}
\begin{document}

\markboth{Dalip Singh Verma and Atul Choudhary}{The  fusion dynamics for a positive Q-value system: $^{27}$Al+$^{45}$Sc using SEDF and role of spin ...}

\catchline{}{}{}{}{}

\title{The  fusion dynamics for a positive Q-value system: $^{27}$Al+$^{45}$Sc using SEDF and role of spin-orbit interaction potential}

\author{Dalip Singh Verma\footnote{dsverma@cuhimachal.ac.in} ~~ and Atul Choudhary }

\address{Department of Physics and Astronomical Science,\\ Central University of Himachal
Pradesh, Dharamshala,\\District Kangra,(H.P)-176215, INDIA\\}
\maketitle
\begin{history}
\received{Day Month Year}
\revised{Day Month Year}
\end{history}
\begin{abstract}
The fusion dynamics for a positive Q-value systems:
$^{27}$Al+$^{45}$Sc, at  near and deep sub-barrier energies has been
investigated using the proximity potentials of Skyrme energy density
formalism in semi classical extended Thomas Fermi approach for
arbitrarily  chosen Skyrme forces: SLy4, SIV, SGII and Proximity77
of Blocki and co-workers. The calculated fusion excitation functions
for the proximity potentials obtained for Skyrme forces mentioned
above and for the Proximity77 have been compared with experimental
data. The proximity potential for Skyrme force  SIV is found to be
the best and is used in the calculations of the quantities like
logarithmic derivative, barriers distributions and $S$-factor.
Further, the role of spin-orbit interaction potential in the fusion
dynamics of this system has been investigated.
\end{abstract}
 \ccode{PACS numbers:24.10.-i, 25.60.Pj, 25.70.-z}
\section{Introduction}{\label{1}}
During the last decade, the  of heavy ion (HI) fusion cross-section
measurements has been extended to deep sub-barrier energies and in
this energy regime the investigation of fusion excitation functions
has revealed new phenomenon like fusion hindrance and enhancement.
The signatures of fusion hindrance, the steep fall in fusion
excitation function, at deep sub-barrier energies with respect to
the standard coupled-channel (CC) calculations was firstly observed
by Jiang and  co-workers \cite{Jiang2002} at National Laboratory,
Argonne, Illinois (USA) in the year 2002. Since then lot of
theoretical and experimental work has been done in this low energy
domain, which is of great interest not only in the field of nuclear
astrophysics, where fusion hindrance can affect the rate of
interstellar reactions like: $^{12}$C+$^{12}$C \cite{Gasques2007},
but it also incorporate with the synthesis of super-heavy elements.

In a quest to understand this unexpected behavior, Jiang {\it et
al.} \cite{Jiang2002} measured and  calculated the fusion
cross-sections using CC calculations and Wong's approximation for
the system $^{60}$Ni+$^{89}$Y at deep sub-barrier energies,
re-analyzed the same for the systems: $^{58}$Ni+$^{58}$Ni \&
$^{90}$Zr+$^{92}$Zr and observed fusion hindrance  for these
systems. These authors proposed that some properties of fused
system, like Q-value (not sufficient alone) as a possible cause of
fusion hindrance. Later Jiang {\it et al.}\cite{JiangBack2006}, has
shown that the fusion hindrance is a general behavior of positive as
well as negative Q-value systems and reproduced the observed
behavior of fusion excitation function by incorporating nuclear
incompressibility in CC calculations. Similarly, the fusion
hindrance is observed\cite{JiangRehm2004} for open-shell system:
$^{64}$Ni+$^{64}$Ni w.r.t CC calculations and the observed data is
reproduced by including nuclear structure inputs. Further, to
explain the fusion hindrance, the double folding potential (which
was too deep for overlapping nuclei) is supplemented with repulsive
core by Misicu and Esbensen \cite{Misicu2007}  along with the
effective nucleon-nucleon interaction ($v_{NN}$),  called shallow
ion-ion potential. In this context, for looking appropriate
potentials, different authors have given different reasons, like
Ichikawa {\it et al.} \cite{Ichikawa2007} has suggested two-step
model of fusion, where the first step is again determined by CC
calculations and second step is determined by using one-dimensional
adiabatic potential barrier and accounts well for fusion hindrance
observed for the systems: $^{64}$Ni+$^{64}$Ni \&
$^{58}$Ni+$^{58}$Ni, Ichikawa {\it et al.} \cite{Ichikawa2009} has
extended the standard CC calculations by considering smooth
transition from sudden to adiabatic states and explained the fusion
hindrance for the systems: $^{16}$O+$^{208}$Pb, $^{64}$Ni+$^{64}$Ni
\&  $^{58}$Ni+$^{58}$Ni, Hagino {\it et al.} \cite{Hagino2003}
fitted the Wood-Saxon potential in CC calculations and concluded
that the large diffuseness is required to predict low energy
behavior of fusion cross-section and Brink {\it et al.} \cite{Brink
2004} suggested that fusion hindrance may be associated with the
overlapping of densities of the two colliding nuclei, where the
description of the potential failed, even when CC effects are
included. Several other reasons for the fusion hindrance has been
given in literature
\cite{Ramamurthy1990,Lin2003,Giraud2004,Seif2004,Sastry2005,Chamon2007,Shilov2012,
Sargsyan2010,Umar2008,Kuzyaki2012,Jiang2013}. In all the
calculations discussed above, the experimental behavior is
reproduced by modifying nuclear potential in different ways.

So, it is of great interest to see whether the  potentials  obtained
in  Skyrme energy density formalism (SEDF) in semi-classical
extended Thomas-Fermi approach (see Ref. \cite{dalip2009}) and
Blocki's proximity potential (Proximity77) \cite{Blocki} are capable
of explaining this enigmatic behavior of fusion excitation function
or not. As the SEDF  has ability to provide different-different
nuclear proximity potentials for different Skyrme forces and is
expressed as the sum of two  different potentials: (i) spin-orbit
density dependent (mainly repulsive) and (ii) spin-orbit density
independent (mainly attractive). Since spin-orbit interaction
potential is mainly repulsive, which is equivalent to the addition
of repulsion to the proximity potential,  therefore it will be
further interesting to see the effects of this interaction potential
in the fusion dynamics for the system: $^{27}$Al+$^{45}$Sc, where
the fusion cross-section has been measured\cite{CL2010} at deep
sub-barrier energies down to about 300 \hbox{nb}, because the
standard CC calculations failed to reproduce the observed data for
this system. Note that the SEDF proximity potentials and Proximity77
are used in Wong's approximation for the  calculation of fusion
cross-section as a function of center of mass energy.

Section 2 describe the methodology for the calculation of interaction potentials, cross-sections, logarithmic derivative, barrier distribution and astrophysical factor. In section 3, calculations and results are discussed and finally section 4 contains the conclusion of the study.

\section{Methodology} The following sub-sections give the calculations detail  for the
Skyrme interaction potential based upon SEDF and Proximity77, fusion cross-section, logarithmic derivative,
barrier distribution and finally the astrophysical factor.
\subsection{Skyrme Energy Density Formalism}
The SEDF, defines the nuclear interaction potential as a function of inter-nuclear separation
$(R)$ as the difference between the energy expectation
value  \textit{E} of the colliding nuclei that are overlapping (at a
finite separation distance $R$) and are completely separated (at
$R=\infty$ ) and is given as,
\begin{eqnarray}\label{e1}
 V(R)&=&
E(R)- E(\infty)
\end{eqnarray}
 where \textit{E} =
 $\int  H(\vec r)d^{3}\vec{r},~ $ is the energy expectation value.
The Skyrme Hamiltonian density (see Refs. \cite{dalip2009,vau1972})
given as,
\begin{eqnarray}\label{e5}
H[~\rho, ~ \tau,~
\vec{J}~]&=&\frac{\hbar^{2}}{2m}\tau_{q}+\frac{1}{2}t_{0}\biggl[\left(1+\frac{x_{0}}{2}\right)\rho^{2
}-\left(x_{0}+\frac{1}{2}\right)\left(\rho_{n}^{2}+\rho_{p}^{2}
\right)\biggr]\nonumber\\
&+&\frac{1}{12}t_{3}\rho^{\alpha_{0}}\biggl[\left(1+\frac{x_{3}}{2}
\right)\rho^{2}-\left(x_{3}+\frac{1}{2}\right) \left(\rho_{n}^{2}+\rho_{p}^{2} \right)\biggr]\nonumber\\
&+&\frac{1}{4}\biggl[t_{1}\left(1+\frac{x_{1}}{2}\right)+t_{2}
\left(1+\frac{x_{2}}{2}\right)\biggr]\rho\tau
-\frac{1}{4}\left[t_{1}\left(x_{1}+\frac{1}{2}\right)-t_{2}\left(x_{2}+\frac{
1}{2}\right)\right]\nonumber\\
&\times&
\left(\rho_{n}\tau_{n}+\rho_{p}\tau_{p}\right)+\frac{1}{16}\biggl[3t_{1}\left(1+\frac{x_{1}}{2}\right)-t_{2}
\left(1+\frac{x_{2}}{2}\right)\biggr]\left(\vec {
\nabla}\rho\right)^{2}\nonumber
\\&-&\frac{1}{16} \biggl[ 3t_ { 1 } \left(x_{1}
+\frac { 1 } { 2 }\right)+ t_{2}\biggl(x_{2}+\frac
{1}{2}\biggr)\biggr]
\left\{\left(\vec{\nabla}\rho_{p}\right)^{2}+\left(\vec{ \nabla }
\rho_{n}\right)^{2}\right\}\nonumber\\
&-&\frac{1}{2}W_{0}\biggl[\rho\vec{\nabla}.\vec{J}+\rho_{n}\vec{
\nabla } .\vec {J}_{n}+\rho_{p}\vec{\nabla}.\vec{J}_{p}\biggr]
\end{eqnarray}
where $\rho=\rho_{p}+\rho_{n},\space
\tau=\tau_{p}+\tau_{n}\space,\vec{J}=\vec{J}_{n} +\vec {J}_ { p }$
are nuclear, kinetic, and spin-orbit densities respectively,
$x_{i},t_{i},\alpha_0, \text{and}$ $W_{0}$ are the Skyrme force
parameters, fitted by different authors, (see e.g. Refs.
\cite{Brack275, Fed})  to obtain ground state properties of the
nuclei.

The kinetic energy density for nucleon $\tau_{q},$ upto second order
of expansion (enough for numerical convergence \cite{Bartel2002}),
in the semi-classical extended Thomas Fermi (ETF) approach of SEDF,
is
\begin{eqnarray}\label{eq12}
\tau_{q}(\vec
r)&=&\frac{3}{5}\left(3\pi^{2}\right)^{2/3}\rho_{q}^{5/3}+\frac{1
}{36}\frac{(\vec{\nabla}
{\rho_{q}})^{2}}{\rho_{q}}+\frac{1}{3}\Delta\rho_{q}+\frac{1}{6}\frac{\vec{
\nabla}{\rho_{q}}.\vec{\nabla}{f_{q}+\rho_{q}}\Delta{f_ { q
}}}{f_{q}} \nonumber
\\&-&\frac{1}{12}\rho_{q}\left(\frac{\vec{\nabla}f_{q}}{f_{q}}\right)^{2}+\frac
{ 1 } {2}\rho_{q} \left(\frac{2m}{\hbar^{2}}\right)^{2} \left(
W_{0}\frac{\vec{\nabla}\left(\rho+\rho_{q}\right)}{2f_{q}}\right)^{2}
\end{eqnarray}
where $f_{q}(\vec r)$ is effective mass form factor, which depend
upon the effective mass term ($3^{rd}$ term of Eq. (\ref{e5})) given
as:
\begin{eqnarray}\label{eq15}
 f_{q}(\vec
r)=\frac{m}{m^{*}({\vec{r}})} &=& 1+\frac { 2m } { \hbar^ { 2 } }
\frac { 1 } { 4 } \biggl\{t_{1}\biggl(1+\frac{x_{1}}{2}\biggr)+
t_{2}\biggl(1+\frac{ x_ { 2 } } { 2 }
\biggr)\biggr\}\nonumber\\&-&\frac{2m}{\hbar^{2}}\frac{1}{4}\biggl\{t_{1}\left(x_{1}
+\frac { 1}{2}\right)- t_ {2}\left(x_{2}+\frac{1}{2}\right)\biggr\}
\rho_{q}(\vec{r})
\end{eqnarray}
 where $q =n ~ or~ p$ for neutron or proton, and  $m^{*}(\vec r)$ is  effective
mass.

The spin $\vec {J}$ being purely a quantal property has no contribution to
the semi-classical functional in
the lowest order, the
second order-contribution is,
\begin{equation}\label{e16}
\vec{J}_{q}(\vec{r})=-\frac{2m}{\hbar^{2}}\left(\frac{1}{2}W_{0}\right)\frac{1}{f_{
q } } \rho_ { q } { \vec\nabla } (\rho+\rho_{q})
\end{equation}
The nuclear interaction potential is calculated in  slab
approximation, for details see  Refs. \cite{dalip2009,dalip2007},
given as,
\begin{eqnarray}\label{e2}
 V_{N}(R)&=&2\pi\bar{R}\int^{\infty}_{s_{0}}e(s)ds,
\end{eqnarray}
 where
$\bar{R}={R_{01}R_{02}}/(R_{01}+R_{02})$ is the mean curvature
radius, $R_{0i}$ are the central radii of interacting nuclei and $e(s)$ is the interaction energy per unit area between
two flat slabs of semi-infinite nuclear matter with surfaces
parallel to the $x-y$ plane and moving in the $z$-direction separated by distance $s$, having a minimum value $s_{0}$, is given
by,
\begin{eqnarray}\label{e3}
\int_{s_0}^{\infty} e(s)ds &=& \int\biggl[ H(\rho,\tau,\vec{J})-\{
H_1(\rho_1,\tau_1,\vec{J}_1)+ H_2(\rho_2
,\tau_2,\vec{J}_2)\}\biggr]dz.
\end{eqnarray}
As $
\tau$ and $\vec{J}$, both are function of
$\rho$, so $H[\rho, \tau, \vec{J}]$ becomes function of $\rho$ only.
Since both spin-orbit density dependent part (repulsive) and
independent part (attractive) behaves differently, so Eq. (\ref{e2})
can be written as,
\begin{equation}\label{eq18}
 V_{N}(R)=V_{P}(R)+V_{J}(R)
\end{equation}
The nuclear density,
$\rho_{i}=\rho_{n_{i}}+\rho_{p_{i}}$, where
$\rho_{n_{i}}=(N_{i}/A_{i})\rho_{i}$ and
$\rho_{p_{i}}=(Z_{i}/A_{i})\rho_{i}$, $i=1,~2$ for two colliding
nuclei \cite{dalip2009}, is the two parameter TF density and in  slab approximation  with temperature dependence included becomes,
 \begin{eqnarray}\label{eq19}
\rho_i(z_i,T)=\rho_{0i}(T)\left[1+\exp\left(\frac{z_i-R_{0i}(T)}{a_{0i}
(T)}\right)\right]^{-1}
 \end{eqnarray}
with $-\infty\leq z\leq \infty~ and~ z_2=R-z_1$, and central density
is,
\begin{eqnarray}\label{eq20}
\rho_{0i}(T)=\frac{3A_i}{4 \pi
R_{0i}^3(T)}\left[1+\frac{\pi^2a_{0i}^2(T)}{R_{0i}^2(T)}\right]^{-1}
\end{eqnarray}
The central radii $R_{0i}(T=0)$, surface thicknesses $a_{i0}(T=0)$
and temperature dependence of these parameters are taken,  from Ref.
\cite{dalip2009}. The Proximity77 is tried here as an alternate to
the SEDF proximity potential, is given as:
\begin{equation}\label{eqq251}
V_{N}=4\pi\bar{R}\gamma b \Phi(s)
\end{equation}
where  $\gamma=0.9517[1-1.7826\{(N-Z)/A\}^{2}]$ MeV fm$^{-2}$ is  nuclear surface energy coefficient, $b=0.99$  fm is nuclear surface width and $\Phi$ is the universal function, which depends upon the separation '{\it s}' between the surfaces of interacting nuclei and is given as,
\[
\Phi(s)=
     \begin{cases}
 -\frac{1}{2}(s-s_{0})^{2}-0.0852(s-s_{0})^{3}; \quad s\le \text{1.2511} \\-3.437\exp\left(-\frac{s_{0}}{0.75}\right);\quad s\ge \text{1.2511}
     \end{cases}
\]
where $s_{0} = 2.54$ is the minimum separation between two flat slabs.
The Coulomb contribution, $V_{C}(R)=k Z_{1}Z_{2}/R$ is added directly to
Eq. (\ref{eq18}) or Eq. (\ref{eqq251}) to obtain the total interaction potential $V_T(R)$, where \textit{k} = 1.44 MeV fm.

To calculate the fusion cross-section as a function of centre of mass-energy ($E_{cm}$), the barrier height ($V_{B}$), barrier position
($R_{B}$) and  barrier curvature ($\hbar
\omega_{0}$) of total interaction potential are used in Wong's formula \cite{Wongs}, is given as
\begin{eqnarray}\label{eq28}
\sigma(E_{cm})=\frac{\hbar\omega_{0}R_{B}^{2}}{2E_{cm}}\ln\left[1+\exp{\left\{
\frac {2\pi}{\hbar\omega_{0}}(E_{cm}-V_{B}) \right\}}\right]
\end{eqnarray}
where $\hbar\omega_{0}$  is obtained in parabolic or inverted
harmonic approximation. Further to amplify the fusion cross-section
at deep sub-barrier energies and near barrier energies the
logarithmic derivative of energy weighted cross-section and barrier
distribution has been calculated respectively. The logarithmic
derivative introduced by Jiang {\it et al. }\cite{Jiang2002} is
given as,
\begin{eqnarray}\label{eq29}
L(E_{cm})&=&d\left[\ln(E_{cm}\sigma)\right]/dE_{cm}=\frac{1}{ \sigma
E_{cm} } \frac { d(E_{cm}\sigma) } { dE_{cm} }.
\end{eqnarray}
The barrier distribution of energy weighted cross-section is \cite{Rowley},
\begin{equation}\label{eq30}
 B(E_{cm})=\frac{d^{2}(E_{cm}\sigma)}{dE_{cm}^{2}}
\end{equation}
An alternative representation of fusion cross-section, frequently
used in nuclear astrophysics to  predict low energy behavior of
nuclear reactions, is astrophysical factor \cite{burbidge} given as
\begin{equation}\label{bc}
 S(E_{cm})= E_{cm}\sigma\exp\{2\pi(\eta-\eta_{0})\}
\end{equation}
 where $\eta (=0.1575Z_{1}Z_{2}\sqrt{\mu/ E_{cm}})$ is the Sommerfield parameter, $Z_{1},
Z_{2}$ are the charges of interacting nuclei and $\mu(=A_1A_2/(A_1+A_2))$ is the reduced mass.  The parameter $\eta_{0}$ is calculated at Coulomb energy $(E_c
=Z_{1}Z_{2}e^{2}/({R_{1}+R_{2})})$. The strong energy dependence of
fusion excitation function depends upon Coulomb barrier,  and  the
$S$-factor remove this dependence by eradicating the Coulomb
component.
\section{Calculations and Results}
First, we have calculated the fusion excitation functions for a
positive Q-value (Q = 9.63 \hbox{MeV}) system: $^{27}$Al+$^{45}$Sc,
using two types of potentials. One  obtained within the
semi-classical extended Thomas-Fermi approach of SEDF for
arbitrarily chosen Skyrme forces SLy4, SIV, SGII and other obtained
from the proximity pocket formula (Proximity77).
\begin{figure}[h!]
\vspace{-0.6 cm} \centering
\includegraphics[ height=0.72\textwidth, clip=true,angle=0,keepaspectratio]{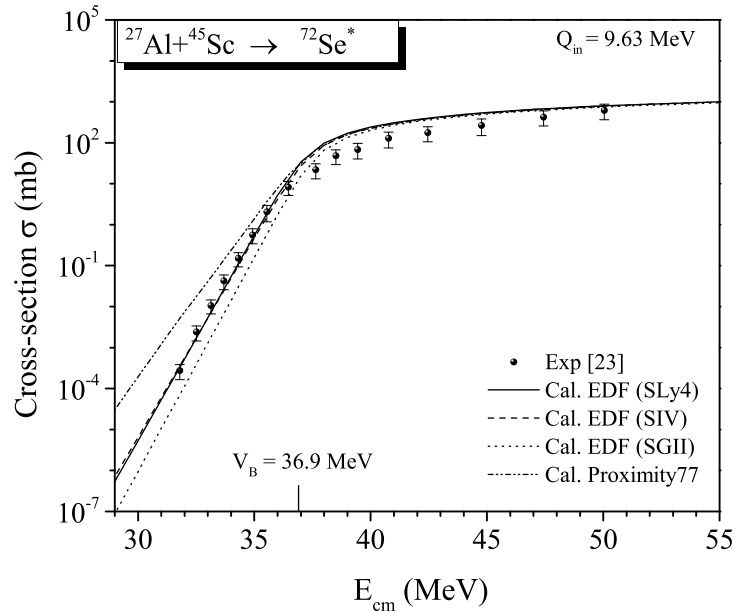}
\caption{A comparison of  the calculated fusion excitation functions
with the proximity potential for arbitrarily chosen Skyrme forces:
SLy4, SIV and SGII and for the fusion excitation function obtained
using Proximity77 with experimental data \cite{CL2010} as a function
of $E_{cm}$.} \label{Fig.1}
\end{figure}
The potential characteristics:- barrier position ($R_B$), barrier
height ($V_B$) and barrier curvature $(\hbar \omega_{0})$ of the two
types of the potentials are used in Wong's formula, Eq.
(\ref{eq28}), to obtain the fusion cross-section as a function of
$E_{cm}$ (i.e. the fusion excitation function). Out of SEDF
proximity  potentials for SLy4, SIV, SGII and Proximity77 used in
the fusion excitation function calculations, the proximity potential
for Skyrme force SIV has been found to be the best (see Fig.
\ref{Fig.1}) and hence is chosen for rest of the calculations. Next,
the role of spin-orbit interaction potential in the fusion dynamics
of the system i.e. in the calculations of the fusion cross-section,
logarithmic derivative, barrier distribution and $S$-factor has been
investigated.

Figure \ref{Fig.1}, shows the comparison of the calculated fusion
excitation functions for the proximity potentials obtained for
Skyrme forces SLy4 (solid line), SIV (dashed line) and SGII (dotted
line) with the experimental data \cite{CL2010} (solid spheres with
error bars). The  dash dot-dot line represent the fusion excitation
function obtained using  Proximity77. It is found from the
comparison of the calculated fusion excitation functions, shown in
Fig. \ref{Fig.1}, with the observed data\cite{CL2010} that  for
below barrier energies the cross-section is best reproduced by the
proximity potential for two forces: SLy4 and SIV. The excitation
functions with the potentials obtained for Skyrme forces SGII and
for the Proximity77 are under-estimating and over-estimating with
respect to the experimental data, respectively. Above  barrier
energies all the potentials are showing almost equivalent results
but among the potentials for SLy4 and SIV Skyrme forces, SIV is
closer to the experimental data and is chosen for rest of the
calculations. As in Ref.\cite{Montagnoli2012} repulsion/repulsive
potential is added to the proximity potential to reproduce the
observed data.
\begin{figure} [h]
\vspace{-0.6 cm}
\centering
\includegraphics[height=0.72\textwidth, clip=true,angle=0,keepaspectratio]{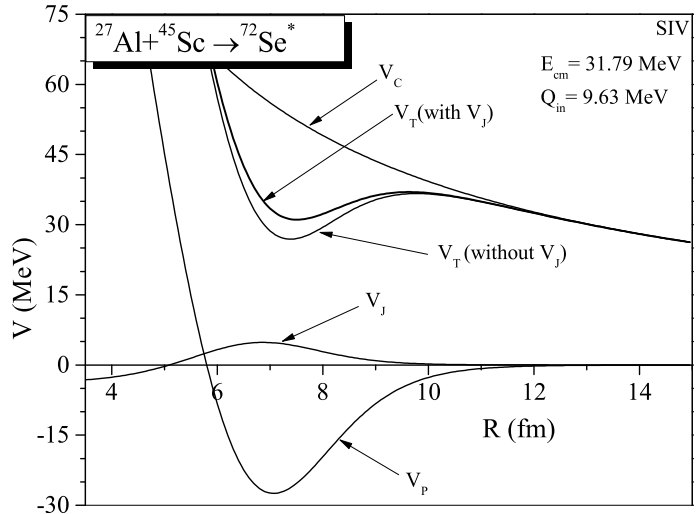}
\vspace{-0.8cm}
\caption{The nuclear proximity potential $V_P$, spin-orbit interaction potential $V_J$, Coulomb interaction potential $V_C$ and the total  interaction potential $V_{T}$  with  and
without spin-orbit interaction potential as a function of inter-nuclear separation R for the Skyrme force SIV.}
\label{Fig.2}
\end{figure}
In SEDF, the nuclear interaction potential has a repulsive part
(spin-orbit density dependent part) in addition to the attractive
part (spin-orbit density independent term) and hence the effect of
latter part of interaction potential on fusion dynamics of
$^{27}$Al+$^{45}$Sc has been studied for the first time and is
discussed below.

Figure \ref{Fig.2} shows different contributing potentials: $V_{P},
V_{J}$ and $V_{C}$ to the total interaction potential $V_{T}$ (with
and without $V_J$) as a function of inter-nuclear distance (R). It
is clear from the Fig. \ref{Fig.2} that the addition of spin-orbit
interaction potential leads to the following minor changes: (i)
shift of potential barrier to the lower position (i.e. $R_B$ changes
from 9.801 \hbox{fm} to 9.671 \hbox{fm}) and (ii) $V_{B}$ changes
from 36.661 \hbox{MeV} to 36.987 \hbox{MeV} (iii) $\hbar \omega_{0}$
from 2.91 \hbox{MeV} to 2.71 \hbox{MeV}.
\begin{figure}[h!]
\vspace{-0.6 cm}
\centering
\includegraphics[height=0.72\textwidth,  clip=true,angle=0,keepaspectratio]{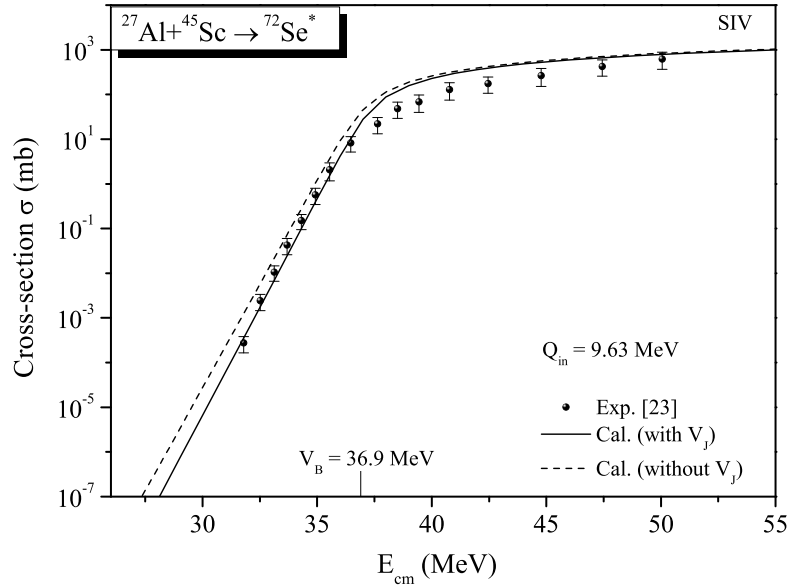}
\caption{A comparison of the calculated fusion excitation function
with and without spin-orbit interaction potential in proximity
potential for Skyrme force SIV  with experimental data\cite{CL2010},
as a function of $E_{cm}$.} \label{Fig.3}
\end{figure}
The corresponding changes in fusion excitation function is shown in
Fig. \ref{Fig.3}, where the fusion excitation function without
$V_{J}$ contribution in total interaction potential (dashed line)
changes to the fusion excitation function (solid line) when $V_{J}$
contribution is included.  It is clear from the Fig. \ref{Fig.3},
that the total interaction potential with spin-orbit interaction
potential is reproducing data nicely. This means in Skyrme energy
density formalism, the addition of spin-orbit interaction potential
that is the repulsive part to the proximity part of the nuclear
potential  enable it to reproduce the experimental data below the
barrier very well. Similar calculations are presented in Ref.
\cite{CL2010} using shallow potential model developed to study the
fusion hindrance in negative Q-value systems, where repulsion is
added to the interaction potential to reproduces the experimental
data.
\begin{figure}[h!]
\centering
\vspace{-0.6 cm}
\includegraphics[height=0.72\textwidth,  clip=true,angle=0, keepaspectratio]{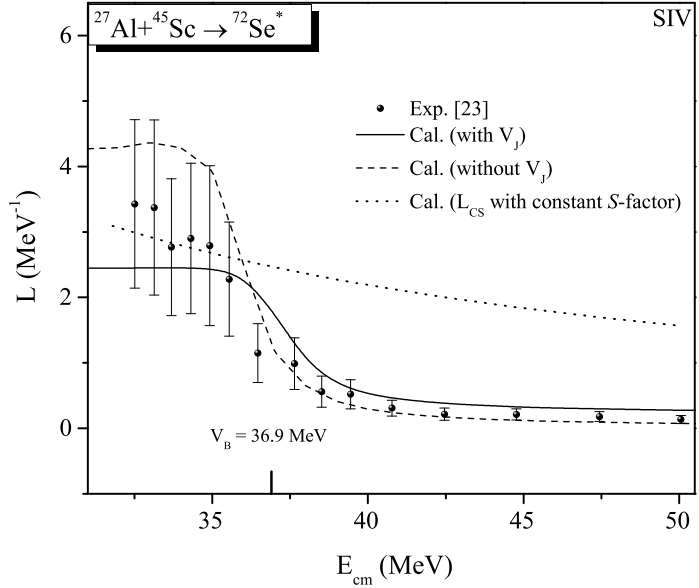}
\caption{The calculated logarithmic derivative of energy weighted
fusion cross-section with and without spin-orbit interaction
potential in proximity potential for Skyrme force SIV and is
compared with the experimental data \cite{CL2010} as a function of
$E_{cm}$.} \label{Fig.4}
\end{figure}

In Fig. \ref{Fig.4}, the logarithmic derivative of energy weighted
fusion cross-section have been shown, the solid line shows the
logarithmic derivative derived from the fusion excitation function
calculated using the total interaction potential including the
spin-orbit interaction term, while the dashed line is for the same
but calculated by excluding spin-orbit interaction potential. Both
these curves develop rapidly with decrease in center of mass energy
near the interaction barrier energies and saturate  further with the
decrease in $E_{cm}$-value, showing that $L$ is independent of
$E_{cm}$. The dotted line is for the logarithmic derivative
corresponding to constant $S$-factor, given as
$L_{CS}=\pi\eta/E_{cm}$, the value corresponding to $E_{cm}$ where
$S$-factor develops maxima. In other words $S$-factor may observe
maxima where $L_{CS}$ intersect with experimental $L$. The
interaction potentials with and without spin-orbit interaction
potential are reproducing the data with in limits of errors.
\begin{figure}[h!]
\centering
\vspace{-0.6 cm}
\includegraphics[height=0.72\textwidth,  clip=true,angle=0, keepaspectratio]{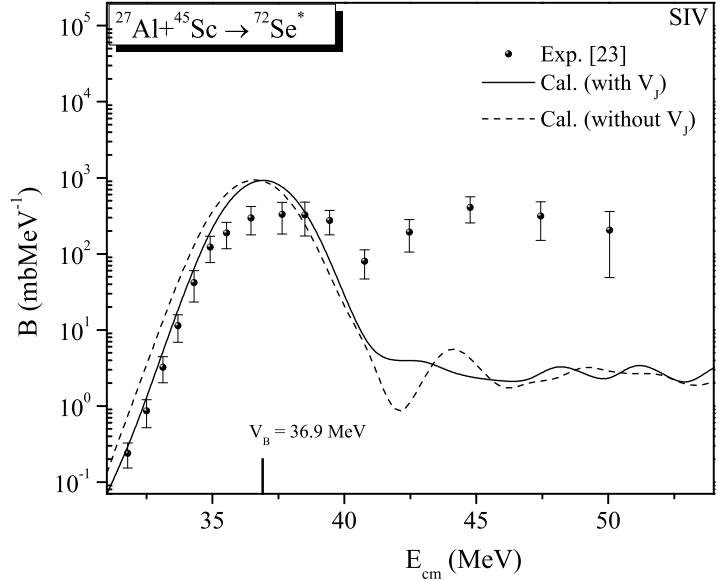}
\caption{The barrier distribution calculated  with  and without
spin-orbit interaction potential in the proximity potential for
Skyrme force SIV  is compared with the experimental
data\cite{CL2010}  as a function of $E_{cm}$.} \label{Fig.4b}
\end{figure}

Figure \ref{Fig.4b} shows a comparison of the  calculated barrier
distribution, which is a magnified view of fusion cross-sections at
energies near the interaction barrier, with the experimental
data\cite{CL2010} (shown by solid spheres with error bars) as a
function of $E_{cm}$. The solid line shows the barrier distribution
when $V_{J}$  term  included in the total interaction potential
while the dashed line is for the same but when $V_{J}$  term is
excluded. Our calculations with  $V_{J}$ term is reproducing the
data nicely near and below barrier energies and is showing an
oscillating behavior for above barrier energies, the oscillations
are stronger for the case when total interaction potential exclude
$V_{J}$ term. Note that the interaction barrier height $V_{B}$ =
36.9 \hbox{MeV} is for the interaction potential including
spin-orbit term.
\begin{figure}[h!]
 \vspace{-0.6cm}
 \centering
\includegraphics[height=0.72\textwidth,  clip=true, angle=0, keepaspectratio]{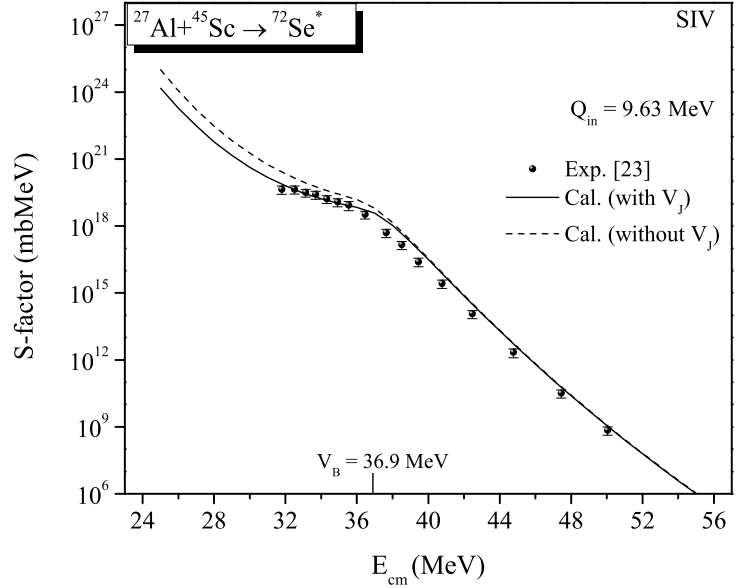}
 \caption{The $S$-factor calculated with and without spin-orbit interaction potential in
 the proximity potential for Skyrme force SIV and is compared with experimental data\cite{CL2010}.}
 \label{Fig.5}
 \end{figure}

Alternate representation to the fusion cross-section is $S$-factor
and is required due to the fact that the fusion cross-section varies
by many order of magnitude below the interaction barrier, here for
this system $^{27}$Al + $^{45}$Sc it is $6$ orders, and at the same
time it removes the dominating influence of the Coulomb barrier
transmission factor that inhibits the broad resonance structure in
fusion excitation function \cite{Erb85}. Figure \ref{Fig.5} shows
comparison of the calculated $S$-factor with experimental
data\cite{CL2010} (solid spheres) where solid line is for the case
when $V_{J}$ is included in the total interaction potential and
dashed line when $V_{J}$ is excluded. For energies above barrier
both cases gives same results while below barrier energies the
interaction potential including the spin-orbit term is nicely
reproducing the experimental results. The $S$-factor increases
rapidly with decrease in center of mass energy and almost saturate
near  the barrier energies and then again increases rapidly with
further decrease in $E_{cm}$. The  broad maxima near deep sub
barrier energies may be an indication for the resonance structure in
the excitation function but the confirmation demands the
experimental data to be available that lower $E_{cm}$. Our
calculations shows  rapid increase in $S$-factor after getting
saturation region and hence is not in favor of resonance structure
in the fusion excitation function for this system.
\section{Conclusion}
We conclude that the interaction potential obtained in SEDF for
Skyrme force SIV is found to be the better than potentials obtained
for SLy4, SGII and  the Proximity77 potential for the fusion
dynamics of $^{27}Al+^{45}Sc$ system. The inclusion of spin-orbit
interaction potential in the proximity potential, which is
equivalent to the addition of repulsion to the nuclear potential,
reproduces the observed data better than when it is excluded.  The
calculated barrier distribution gives comparatively stronger
oscillations  for interaction potential excluding spin-orbit term
for energies above barrier. The clear maxima in $S$-factor is not
found, for deep sub-barrier energies, in our calculations as well as
in the experimental data which indicates the absence of structure in
fusion excitation function. Our calculations shows the strong energy
dependence of astrophysical factor for far below the barrier
energies and hence more experimental data is required at these
energies to see the behavior of fusion excitation functions for this
system.

\end{document}